\documentclass[conference]{IEEEtran}
\usepackage{balance}

\usepackage{tikz}
\usepackage{amsmath}

\usepackage[autostyle]{csquotes}
\usepackage[acronym]{glossaries}
\setacronymstyle{long-short}
\newacronym{iot}{IoT}{Internet of Things}
\newacronym{sd}{SD}{Standard Deviation}
\newacronym{ap}{AP}{Access Point}
\newacronym{XGBoost}{XGBoost}{Extreme Gradient Boosting}
\newacronym{dt}{DT}{Decision Tree}
\newacronym{rf}{RF}{Random Forest}
\newacronym{lgbm}{LGBM}{Light Gradient Boosted Machine}
\newacronym{ecdf}{ECDF}{Empirical Cumulative Distribution Function}
\newacronym{cdf}{CDF}{Cumulative Distribution Function}
\newacronym{nic}{NIC}{Network Interface Card}
\newacronym{qos}{QoS}{Quality of Service}
\newacronym{cu}{CU}{Channel Utilization}
\newacronym{ml}{ML}{Machine Learning}
\newacronym{rtt}{RTT}{Round-Trip Time}
\newacronym{ms}{ms}{millisecond}
\newacronym{rpi}{RPi}{Raspberry Pi}
\newacronym{dma}{DMA}{Direct Memory Access}
\newacronym{pdf}{PDF}{Probability Density Function}
\newacronym{RSS}{RSS}{Residual Sum of Squares}

\begin{document}

%\title{Accurate Identification of IoT Devices in the Presence of Wireless Channel Dynamics}

\title{Leveraging Machine Learning for Accurate IoT Device Identification in Dynamic Wireless Contexts}
\twocolumn

\author{\IEEEauthorblockN{Bhagyashri Tushir, Vikram K Ramanna, Yuhong Liu, Behnam Dezfouli}
\IEEEauthorblockA{\small Internet of Things Research Lab, Department of Computer Science and Engineering, Santa Clara University, USA}
\texttt{\small\{btushir, vramanna, yhliu, bdezfouli\}@scu.edu\quad}
}

\maketitle
%\pagenumbering{arabic}

\thispagestyle{plain}
\pagestyle{plain}

\begin{abstract}

Identifying IoT devices is crucial for network monitoring, security enforcement, and inventory tracking. However, most existing identification methods rely on deep packet inspection, which raises privacy concerns and adds computational complexity. More importantly, existing works overlook the impact of wireless channel dynamics on the accuracy of layer-2 features, thereby limiting their effectiveness in real-world scenarios. In this work, we define and use the latency of specific probe-response packet exchanges, referred to as "device latency," as the main feature for device identification. Additionally, we reveal the critical impact of wireless channel dynamics on the accuracy of device identification based on device latency. Specifically, this work introduces "accumulation score" as a novel approach to capturing fine-grained channel dynamics and their impact on device latency when training machine learning models. We implement the proposed methods and measure the accuracy and overhead of device identification in real-world scenarios. The results confirm that by incorporating the accumulation score for balanced data collection and training machine learning algorithms, we achieve an F1 score of over 97\% for device identification, even amidst wireless channel dynamics, a significant improvement over the 75\% F1 score achieved by disregarding the impact of channel dynamics on data collection and device latency.

\end{abstract}

% \begin{IEEEkeywords}
% Security, Wi-Fi, Channel Utilization, Machine Learning, Privacy, Network Management
% \end{IEEEkeywords}
% \IEEEpeerreviewmaketitle

\glsresetall

\section{Introduction} 
\label{intro}

The rapid expansion of \gls{iot} devices is remarkable, with projections indicating a rise to over 29 billion devices by 2027.
Wi-Fi plays a significant role in this \gls{iot} revolution, being the backbone for 31\% of \gls{iot} devices' connectivity~\cite{num_iot_devices}. 
In 2022, shipments of Wi-Fi-enabled IoT devices constituted 37\% of the total market, and this figure is expected to surpass 40\% by the year 2027~\cite{Wifi_by_the_num}.

The growth in the number and variety of \gls{iot} devices underscores the critical need for precise device identification.
Effective \gls{iot} device identification empowers network middleboxes and appliances (such as wireless \glspl{ap}, switches, and network controllers) to enhance the management and security of connected devices~\cite{santos2018efficient,tushir2020quantitative}. 
For example, to enhance security, micro-segmentation strategies can be employed to segregate devices based on their functions and security requirements~\cite{osman2020transparent}. 
Additionally, should a device exhibit unusual traffic patterns, indicative of potential security threats~\cite{sivanathan2020detecting}, it can be temporarily isolated for investigation.
Moreover, accurate identification of \gls{iot} devices enables fine-tuning connectivity settings, tailored to the specific needs of each device~\cite{marchal2019audi}. 
For instance, the \gls{ap} can be configured to guarantee the bandwidth of each device based on its type of service~\cite{chen2021predictable}.
This customization not only prioritizes devices with higher importance or specific latency demands, but also significantly improves the overall user experience.
Beyond operational efficiency, \gls{iot} device identification offers valuable insights into device usage, revealing patterns and behaviors instrumental in enhancing existing services or inspiring new product developments. 
Proactive monitoring of devices' performance and status through identification also aids in early detection of potential malfunctions, allowing for timely interventions to prevent failures.

% chowdhury2020network: Network traffic analysis based iot device identification
The conventional techniques for identifying \gls{iot} devices, primarily based on IP and MAC addresses, are  inadequate due to their limited applicability and susceptibility to security threats such as spoofing~\cite{chowdhury2020network}. 
For instance, while MAC addresses can be used to determine the manufacturer of the device, many manufacturers produce a variety of device types (smartphones, laptops, and \gls{iot} devices), so the MAC address alone is not sufficient to precisely determine the type of device.
Additionally, MAC addresses can be spoofed or even intentionally randomized by the device manufacturer to avoid tracking.
Hence, more robust approaches are required for device identification.
To this end, various works have leveraged the unique traffic patterns of \gls{iot} devices~\cite{sivanathan2018classifying, santos2018efficient, thangavelu2018deft, ammar2020autonomous, miettinen2017iot, pinheiro2019identifying, gordon2021securing, gordon2021efficient, aksoy2019automated}. 
However, these studies have some key limitations.
{Firstly}, their proposed features rely on specific network configurations and traffic conditions, limiting their applicability over time and across different deployment environments. 
For instance, while features such as packet length and packet rate are impacted by the dynamics of traffic patterns, wireless channel, and the number of devices accessing the wireless channel, such impacts have been overlooked~\cite{sivanathan2018classifying,pinheiro2019identifying}.
{Secondly}, the process of extracting features at various levels of the protocol stack (including attributes like destination DNS queries, IP addresses, and port numbers) not only risks compromising user privacy due to the sensitivity of the information involved, but also results in increased computational complexity and memory demands~\cite{sivanathan2018classifying}.
{Thirdly}, existing approaches involve collecting training data over the period of one or several weeks, causing a long delay before the data can be used for device identifications~\cite{sivanathan2018classifying}.

In light of the identified challenges, this paper is guided by two key design considerations:
(i) given the sensitivity of IoT devices (such as smart homes where personal data is handled within the privacy of individual homes), it is imperative to identify these devices in a manner that safeguards users' personal information;
(ii) the dynamic nature of deployment environments, characterized by evolving traffic patterns of devices and fluctuating wireless channel properties.
Addressing these critical concerns, this paper makes the following contributions towards more reliable identification of Wi-Fi (802.11) IoT devices.
We introduce measurement techniques and features based on device response times to TCP and UDP probe packets, enabling the fingerprinting of various \gls{iot} device types.
While latency features are useful for privacy-preserving device fingerprinting, our analysis reveals that such features are significantly influenced by wireless channel dynamics, limiting their stability and accuracy under various wireless channel conditions.
Although \gls{cu} is a common metric for assessing wireless channel dynamics, its granularity and reliability are hampered by inherent short-term dynamicity and hardware limitations. 
To overcome these limitations, we introduce \textit{accumulation score} as a robust metric for instantaneous measurement of channel dynamics and their influence on device latency. 
%This approach provides a more reliable metric for understanding and adapting to changing network conditions.
This metric provides a more reliable approach for understanding and adapting to changing network conditions.
We implement the proposed methods and build a testbed to measure identification accuracy and system's overhead in real-world settings.
Our empirical evaluations demonstrate that device identification methodologies that fail to account for channel dynamics yield accuracy rates ranging from 25\% to 75\%. 
In contrast, by incorporating the accumulation score into our data collection and \gls{ml} algorithm training processes, we achieve identification accuracy exceeding 97\% in the presence of diverse wireless channel conditions.
We will also present training and inference overhead analysis of the \gls{ml} algorithms when used on a residential \gls{ap}.

The rest of the paper is organized as follows. 
In Section \ref{prpo_feature}, we study the opportunities and challenges of using device latency for device identification.
In Section \ref{account_for_cu}, we present accumulation score and its importance for device identification.
We present empirical performance evaluation of device identification accuracy and overhead of \gls{ml} algorithms in Section \ref{result}.
We overview related work in Section \ref{realted_work} and conclude the paper in Section \ref{conc}.

\section{Significance of Device Latency}  
\label{prpo_feature}

In this section, we present the motivation behind considering \textit{device latency} as a feature for \gls{iot} device (hereinafter referred to as \textit{device}) identification. 
We also identify the challenges of measuring this feature and discuss our proposed methodology.
Finally, we illustrate how device latency varies among devices across different \gls{cu} ranges.

\subsection{Motivation for Utilizing Device Latency} 
\label{mot_dl}

\begin{figure}[!t]
\centering
\includegraphics[width=1\linewidth]{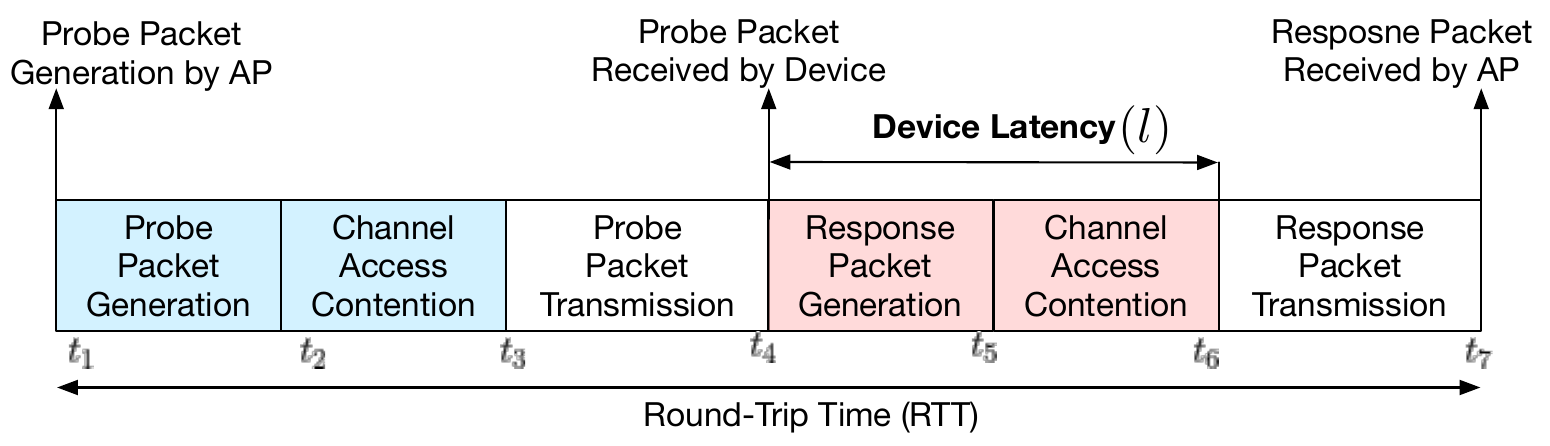}
\caption{Device latency ($l$) is defined as the interval between $t_4$ to $t_6$.
Round-Trip Time (RTT) is defined as the interval between crafting a probe packet by the \gls{ap} until the reception of the response packet (generated by the device) by the \gls{ap}.
Extracting device latency ($t_4$ to $t_6$) from RTT ($t_1$ to $t_7$) is challenging due to the variability of $t_1$ to $t_4$.
}
\label{latency_time_interval}
\end{figure}

In this work, we adopt \textit{device latency} (denoted as $l$) as the primary feature for device identification in Wi-Fi networks.
Device latency is defined as follows: once a packet is received by a device, how long it takes for the device to process the packet, generate a response, and start the transmission of the response packet.
In this paper, we refer to the packets used to measure device latency as \textit{probe packets}.
Referring to Figure~\ref{latency_time_interval}, $t_{4}$ denotes the time instance at which a probe packet has been completely received by the device, and device latency is the interval between $t_{4}$ to $t_{6}$.
Specifically, device latency comprises the following two components: response generation and channel contention, as illustrated in Figure~\ref{latency_time_interval}, corresponding to time intervals $t_{4}$ to $t_{5}$ and $t_{5}$ to $t_{6}$, respectively.
Several factors contribute to the variability of these latency components, including the packet type, the device's hardware characteristics such as processor type and memory, the type of wireless \gls{nic} employed, the \gls{nic} driver in use, the specific operating system and the network stack implemented on the device. 
These attributes collectively influence how a device handles latency-related tasks.
For instance, the processor, \gls{nic} driver, operating system, and network stack impact incoming and outgoing packet processing, while the wireless \gls{nic} and its driver determine how a device contends for channel access before transmitting each packet.
This interrelationship between device latency and device-specific attributes introduces an interesting notion: leveraging device latency as a discriminating feature to differentiate between various devices.

\subsection{Feature Measurement Methodology}
\label{feature_measu}

The precise measurement of device latency is essential for accurate device identification. 
One potential method for measuring device latency is using \gls{rtt}, which involves generating and transmitting probe packets from an \gls{ap} to a device, followed by measuring the \gls{rtt} upon receiving responses.
This process involves various time intervals illustrated in Figure~\ref{latency_time_interval}. 
Specifically, the \gls{rtt} includes: 
(i) packet generation interval, from $t_1$ to $t_2$, which represents the time taken by the \gls{ap} to craft a probe packet, 
(ii) channel access contention by the \gls{ap} during $t_2$ to $t_3$ to send the probe packet,
(iii) actual transmission of probe packet from $t_3$ to $t_4$,
(iv) packet processing and response generation during $t_4$ to $t_5$ by the device, 
(v) channel access contention by the device during $t_5$ to $t_6$ to send the response packet,
and (iv) actual transmission of response packet during $t_6$ to $t_7$.
Among these delay components, the time interval from $t_{1}$ to $t_{4}$ does not represent device latency; therefore, the variability of this time interval can negatively affect device latency measurement accuracy, impacting the effectiveness of device identification.

\begin{figure}[!t]
\centering
\includegraphics[width=0.83\linewidth]{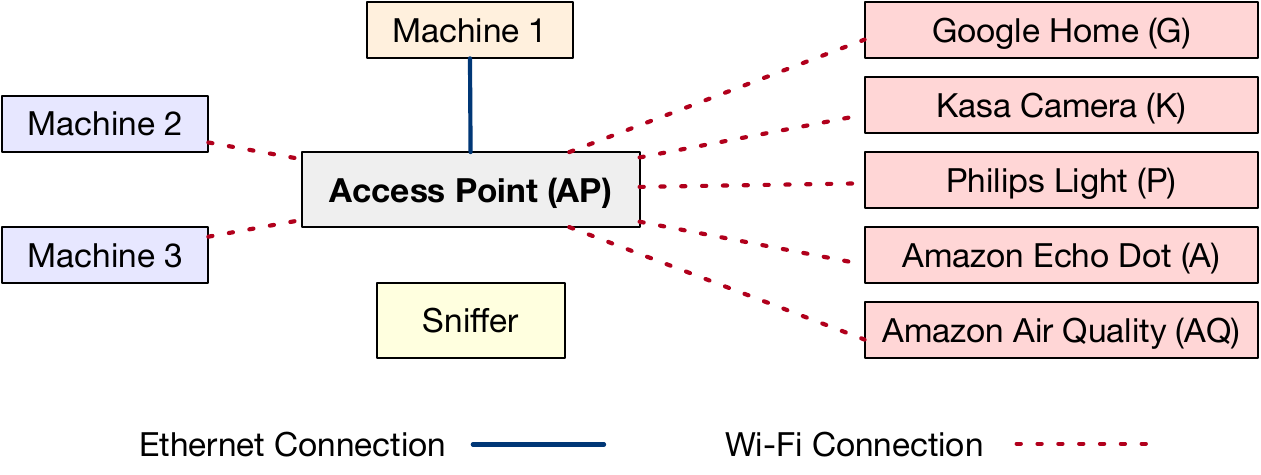}
\caption{The testbed components including machines used for background traffic generation, IoT devices, and sniffer. 
Note that various experiments of this work utilize subsets of these components, depending on the specific objectives and requirements of each study.
}
\label{packet_switch_dc}
\end{figure}

To characterize the impact of the time interval $t_{1}$ to $t_{4}$ and evaluate the feasibility and accuracy of extracting device latency from \gls{rtt}, we measure the variability of $t_{1}$ to $t_{4}$.
To this end, we set up a testbed, as demonstrated in Figure~\ref{packet_switch_dc}, which includes an \gls{ap}, a packet sniffer, and three Linux machines denoted as Machine 1, 2, and 3.
Machine 1 is connected to the \gls{ap}'s Ethernet port, and Machine 2 and Machine 3 are connected to the \gls{ap} wirelessly.
It should be noted that the IoT devices depicted in Figure~\ref{packet_switch_dc} are not utilized in this particular experiment; rather, they are employed in the experiments detailed in subsequent sections.
We craft and transmit TCP-SYN packets as probe packets from the \gls{ap} to Machine 3 every 1 \gls{ms} and measure deviations from this baseline value by capturing the packets via a packet sniffer.
Also, we introduce different packet-switching loads on the \gls{ap} to evaluate their impact on packet generation and channel contention duration.
To this end, Machine 1 transmits traffic, employing the {\fontfamily{pcr}\selectfont iperf} tool, towards Machine 2.
With this setup, we generate network traffic at two distinct rates, namely 25 Mbps and 50 Mbps, effectively simulating different packet-switching loads on the \gls{ap}, while the \gls{ap} continues to transmit probe packets.

\begin{figure}[!t]
\centering
\includegraphics[width=1\linewidth]{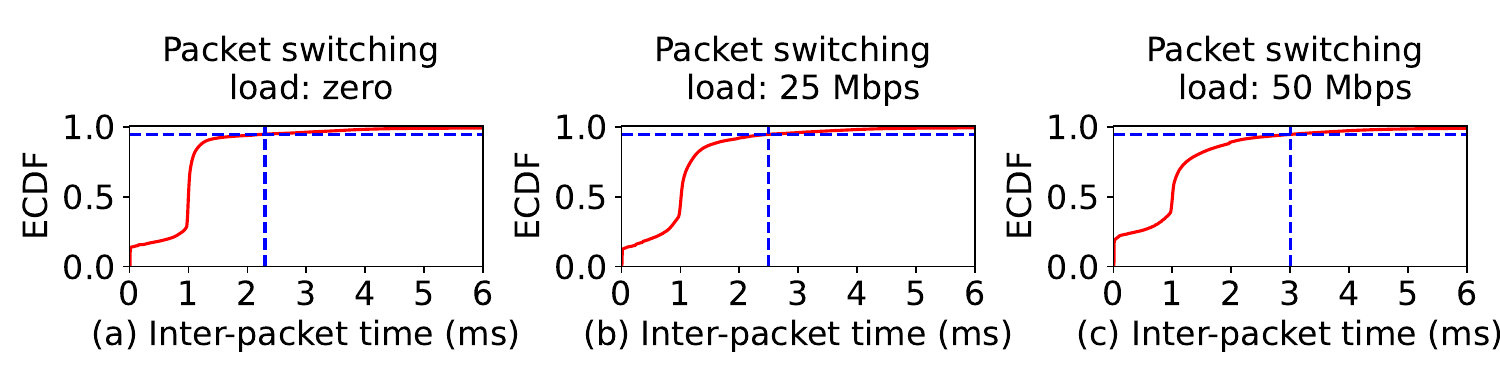}
\caption{Inter-packet intervals when the \gls{ap} is sending a probe packet every 1 ms. The horizontal dashed line represents the 95th percentile, and the vertical line indicates the point where the inter-packet time intersects with the 95th percentile.
}
\label{feature_meas}
\end{figure}

Figures~\ref{feature_meas} (a) through (c) demonstrate the intervals between probe packets sent by the \gls{ap} while the \gls{ap} is under various packet switching loads.
It is evident from these results that inter-packet intervals deviate from the desired 1 ms interval.
Also, as the packet switching load of the \gls{ap} increases, the variations and mean of inter-packet intervals increase as well.
For instance, while 95\% of the inter-packet time is smaller than 2.2 ms in Figure~\ref{feature_meas}(a), this value increases to 3 ms in 
Figure~\ref{feature_meas}(c).
This behavior is attributed to the higher processing load and queuing delay of the \gls{ap}.
Therefore, in a real-world environment where the packet switching load of the \gls{ap} is dynamically changing, the intervals between consecutive probe packets are highly affected.
Hence, this approach is not a reliable method to measure device latency.

We propose an alternative method to address the above challenges.
Instead of relying on the \gls{rtt} method to measure the time intervals encompassing $t_{1}$ to $t_{7}$, we focus on the interval from $t_{4}$ to $t_{7}$ and subtract $t_6$ to $t_7$ to extract device latency.
To this end, we employ a packet sniffer to capture probe packets and the corresponding response packets. 
A program then parses the captured data and correlates probe and response packets for each device. Detailed packet information, such as packet type and duration, can be extracted from the captured data to compute the proposed features.
For instance, if the probe packet is a TCP-SYN sent to a closed port, the packet parser matches the TCP-SYN with a TCP-RST packet by correlating their sequence number fields.
Once a pair of packets (probe and response) is found, the packet parser extracts $t_4$ as the time of receiving the probe packet and $t_7$ as the time of receiving the response packet.
Then, the duration of the response packet ($t_7 - t_6$) is subtracted from $t_7 - t_4$ to extract device latency.

It is worth noting that the packet capture process does not necessarily require a sniffer in addition to the \gls{ap}.
For example, many modern \glspl{ap} provide an additional \gls{nic} that can be used for capturing packets on the desired channel.
Alternatively, the exact transmission and reception time of packets by an interface can be monitored by kernel modification or using eBPF hooks in the driver.
From the memory utilization perspective, considering that Wi-Fi \glspl{ap} typically have limited storage capacity, the proposed approach is designed to optimize storage use. 
For example, the Netgear WAX218 has 256 MB of flash memory and 512 MB of RAM.
The parser program processes captured packet headers in real-time and selectively stores only essential information from the packets of interest.
Specifically, when a probe packet is sent, only specific details of this packet, such as its type (e.g., TCP-SYN), sender and receiver MAC addresses, and its transmission's start and end time, are stored. The parser then enters a state of readiness to identify a corresponding response packet for the probe packet.
Upon detecting the response packet, the program stores information related to the pair of probe-response packets.
It is also essential to note that the packet parser only stores packet latency information and probe packet type for each device.
Therefore, no user-sensitive information (such as destination IP addresses) is stored and used by the identification algorithm.
This approach ensures that the proposed method is privacy-preserving.

\subsection{Device Latency versus Device Type}  
\label{mot_dl}

To empirically validate the significance of device latency in two essential aspects—namely, (i) its role in facilitating device identification and (ii) its sensitivity to variations in \gls{cu} intensity—we conduct preliminary experiments within a smart home testbed. 
%based on the testbed explained in the previous section.
As Figure~\ref{packet_switch_dc} shows, the testbed includes off-the-shelf devices, each denoted by the notation enclosed in parentheses: Google home (G), Amazon echo dot (A), Kasa camera (K), Philips light bulb (P), and Amazon air quality monitor (AQ).
All of these devices are connected to the \gls{ap}.
The \gls{ap} monitors \gls{cu} intensity using the {\fontfamily{pcr}\selectfont ethtool} utility, which captures measurements at 10 \gls{ms} intervals (shortest possible interval) to determine the percentage of time the channel remains occupied during each interval~\cite{sheth2021monfi}.
To evaluate the impact of \gls{cu} on device latency, we adjust \gls{cu} intensity by varying the data rates of the flows being exchanged between Machine 1 and Machine 2 and Machine 3 in Figure~\ref{packet_switch_dc}.
We use {\fontfamily{pcr}\selectfont iperf} on these machines to regulate \gls{cu} intensity, spanning the range from 0\% to 100\%.
To elicit responses from the devices, the \gls{ap} transmits one low-payload (0 byte) TCP-SYN probe packet per second to a closed port of each device, anticipating a corresponding TCP-RST response packet.

\begin{figure*}[!t]
\centering
\includegraphics[width=1\linewidth]{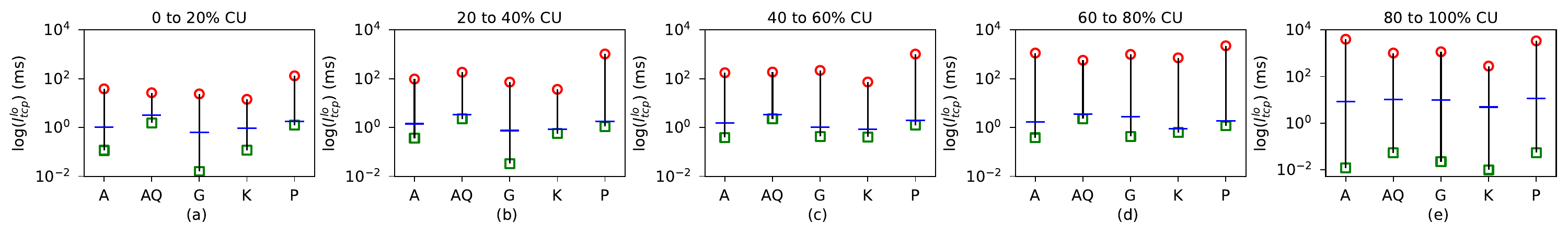}
\caption{Device latency ($l_{tcp}^{lo}$) for various devices and \gls{cu} ranges. 
Results are collected when using low-payload (0 B) TCP-SYN probe packets.
Circles, squares, and horizontal lines represent maximum, minimum, and median values, respectively.
}
\label{device_rtt}
\end{figure*}

Figures~\ref{device_rtt} (a) through (e) present statistical metrics encompassing the median, minimum, and maximum device latency for various devices and \gls{cu} ranges.
Device latency is denoted as $l_{tcp}^{lo}$ because of the use of TCP-SYN probe packets with 0-byte (low) payload size.
The statistical metrics are shown for different ranges of \gls{cu} intensity, corresponding to the five sub-figures.
Three primary observations can be made from Figure~\ref{device_rtt}. 
Firstly, within each \gls{cu} intensity range, the median device latency varies for each device. 
For instance, in Figure~\ref{device_rtt}(a), the air quality monitor (AQ) shows the highest median device latency at 3.27 \gls{ms}, while Google home (G) exhibits the shortest median latency at 0.62 \gls{ms}. 
Secondly, significant differences exist in the minimum and maximum device latency values across all devices, with some overlapping within these metrics.
These differences stem from variations in device hardware and software characteristics. %as elaborated in section \ref{mot_dl}.
Thirdly, when comparing the figures, we observe a simultaneous shift in median, minimum, and maximum device latency with changes in \gls{cu} intensity across all devices. 
This relationship between \gls{cu} intensity and device latency highlights the significance of considering both factors when addressing device identification.

Based on these findings, it becomes evident that leveraging device latency as the key feature for device identification has a significant potential. 
However, relying solely on 0-byte TCP-SYN probe packets may be insufficient for accurate device identification.
With these insights, we introduce four distinct probe packet types to collect device latency data: low-payload (0 B) TCP-SYN, high-payload (1400 B) TCP-SYN, low-payload (0 B) UDP, and high-payload (1400 B) UDP. 
The respective device latency associated with these packet types are labeled as $l_{tcp}^{lo}$, $l_{tcp}^{h}$, $l_{udp}^{lo}$, and $l_{udp}^{h}$.
The motivation for utilizing different payload sizes lies in the fact that larger payloads have an evident impact on various essential stages within the packet processing workflow, such as memory allocation and deallocation for packets, and \gls{dma} of packets from the \gls{nic} to the driver's receiver buffer.
Therefore, exploring how device latency varies across these packet types and sizes yields insights into the latency variation among devices, enabling us to differentiate devices effectively. 
Note that the \gls{ap} sends all these probe packets to the closed ports of the devices. 
Responses to UDP and TCP-SYN probes are ICMP destination unreachable packets and TCP-RST packets, respectively.

\section{Accounting for the Impact of Channel Utilization on Device Latency} 
\label{account_for_cu}

Some wireless \gls{nic}s are capable of measuring and reporting \gls{cu} by dividing channel activity time by a reference time period. 
For instance, Atheros drivers in \gls{nic}s measure \gls{cu} every 10 ms~\cite{sheth2021monfi,sheth2021flip}.\footnote{Tools such as \texttt{ethtool} can be used to retrieve \gls{cu} measurements from the driver.}
Considering that device latency can often be under 10 \gls{ms}, as shown in Figure~\ref{device_rtt}, and given the dynamic nature of \gls{cu}, the coarse granularity of the \gls{cu} measurements performed by these drivers is insufficient to accurately quantify the impact of \gls{cu} on device latency. 
This limitation is increasingly significant in light of the higher physical layer rates introduced by new standards like 802.11be and the use of faster processors on IoT devices.
Therefore, we propose a novel approach to estimate the intensity of \gls{cu} and its impact on each probe-response exchange. 
This method is designed to offer a more granular and accurate analysis, keeping pace with the rapid advancements in wireless communication technologies.

\subsection{Accumulation Score}

\begin{figure}[!t]
\centering
\includegraphics[width=1\linewidth]{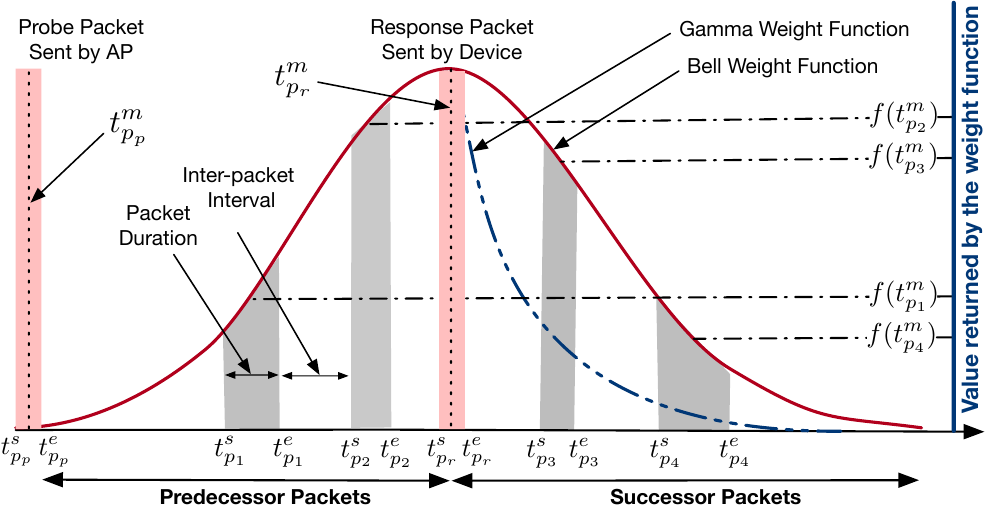}
\caption{A sample probe-response packet exchange with Predecessor and Successor packets.
Accumulation score calculation associated with each device latency value uses three major parameters:
the duration of Predecessor and Successor packets, inter-packet intervals, and value obtained from the projection of packets' midpoints onto a Bell or Gamma curve.
}
\label{as}
\end{figure}

In this section, we present the formulation of the \textit{accumulation score}, a novel metric to measure instantaneous \gls{cu} and its impact on device latency. 
We use the notation $p_{i}$ to denote a generic packet. 
Specifically, probe and response packets are represented as $p_{p}$ and $p_{r}$, respectively.
For each packet $p_i$, we denote its start and end times as $t_{p_{i}}^{s}$ and $t_{p_{i}}^{e}$.
We begin by introducing two essential concepts employed in the accumulation score formulation:
(i) \textit{Predecessor Packets}
and
(ii) \textit{Successor Packets}.
Figure~\ref{as} illustrates these concepts in the context of a probe-response exchange.
\textit{Predecessor Packets} are defined as the packets sent during the time interval between the transmission of the probe packet and the transmission of the response packet. In Figure~\ref{as}, packets $p_{1}$ and $p_{2}$ are examples of Predecessor Packets. 
The occurrence of these packets indicates competition for channel access, potentially increasing the \gls{cu} and resulting in increased device latency, particularly when these packets are transmitted close to the transmission of response packet.
On the other hand, \textit{Successor Packets} refer to the packets transmitted after the response packet, up until a specific time instance that will be further detailed later in this section. Packets $p_{3}$ and $p_{4}$ in Figure~\ref{as} are examples of Successor Packets. 
The rationale for considering Successor Packets lies in their potential impact on device latency, especially when a series of packets immediately follows the response packet. 
A cluster of such packets can suggest possible contention at the time the device was trying to transmit the response packet, thereby influencing device latency.

In evaluating the impact of \gls{cu} on device latency, we focus on certain characteristics of Predecessor and Successor packets. 
These include packet duration, mid-packet points, and inter-packet intervals. 
The duration of a packet, indicating the length of time it occupies the communication channel, is defined as $d_{p_{i}} = t_{p_{i}}^{e} - t_{p_{i}}^{s}$. 
Longer packet durations lead to extended channel occupancy, thereby increasing \gls{cu} intensity and, consequently, device latency.
The midpoint of a packet's transmission, defined as $t^{m}_{p_{i}} = t_{p_{i}}^{s} + \frac{t_{p_{i}}^{e} - t_{p_{i}}^{s}}{2}$, represents the time at the center of the packet transmission.
The inter-packet interval is another critical factor that is defined as $v_{p_{i-1}, p_{i}} = t_{p_{i}}^{s} - t_{p_{i-1}}^{e}$ and refers to the interval between packets $p_{i-1}$ and $p_{i}$.
Shorter inter-packet intervals typically signify higher \gls{cu}, increased contention for channel access, and longer device latency.

To account for the impact of each Predecessor or Successor packet on \gls{cu}, we consider the relative distance (time difference) of each packet from the response packet ($t_{p_{r}}$).
As discussed earlier, the closer a packet is to its corresponding response packet, the higher its impact on device latency.
To reflect this relationship, we define and use \textit{weight functions} that reach their peak at the time of response packet transmission ($t_{p_{r}}^{m}$).
Figure~\ref{as} demonstrates two such weight functions, called the Bell and Gamma curves.
The Bell curve has a mathematical form similar to the \gls{pdf} of the normal distribution and is expressed as $f(x; \mu, \sigma) = \frac{1}{\sigma\sqrt{2\pi}} e^{-\frac{1}{2}\left(\frac{x-\mu}{\sigma}\right)^2}$, where $x$ is the variable of interest, $\mu$ is the mean, and $\sigma$ is the standard deviation.
Similarly, the Gamma curve is expressed as $f(x; \alpha, \beta) = \frac{1}{\beta^\alpha} x^{\alpha-1} e^{-\frac{x}{\beta}}$, where $\alpha$ and $\beta$ are parameters that control the shape and scale of the function.
Note that the mean of both functions is at $t_{p_{r}}^{m}$, the middle point of the response packet.
A gamma curve considers both distance and precedence, meaning that for a Predecessor and a Successor packet both equidistant from $t_{p_{r}}^{m}$, a higher weight is given to the Predecessor packet.
In contrast, the Bell curve takes into account distance only when assigning weight.
It is important to note that these weight functions assign relative weights to each packet, reflecting their contribution to \gls{cu} intensity within a probe-response exchange rather than modeling the actual packet distribution.

To integrate the above metrics and evaluate the impact of Predecessor and Successor packets on each probe-response exchange, we introduce the \textit{accumulation score} metric.
Assume the number of Predecessor and Successor packets may vary for each probe-response exchange, denoted as $n$, with values ranging from 0 to $i$. 
Accumulation score is defined as follows:

\begin{equation} \label{eq:as}
    \mathcal{A}_{j} = 
    \begin{cases}
    0  & \text{ if }  n =0 \\ 
    d_{p_{1}} \times \sigma \times f(t^{m}_{p_{1}})  & \text{ if } n = 1\\ 
    d_{p_{1}} \times \sigma \times f(t^{m}_{p_{1}})  + & \text{ if } n > 1 
    \\
    \indent \sum_{i=2}^{n} (d_{p_{i}} \times \frac{1}{v_{p_{i-1}, p_{i}}/\sigma} \times f(t^{m}_{p_{i}}) ) 
    \end{cases}
\end{equation}

If $n = 0$, we assign a zero value to the accumulation score.
When $n = 1$, since there is no inter-packet interval, we normalize the duration of the packet by multiplying it by the standard deviation of the curve, denoted as $\sigma$.
When selecting a weight function, we select function parameters that maximize the correlation between accumulation score and device latency.
After normalizing the packet duration, the value is multiplied by the value obtained from the weight function.
If $n > 1$, we further consider the impact of the inter-packet interval on the accumulation score for all subsequent packets beyond the first, ensuring that the influence of each packet on the score intensifies as their inter-packet intervals decrease.
When normalizing the duration of a packet, we calculate the ratio of the inter-packet interval between the current packet and the subsequent one to the standard deviation ($\sigma$) of the weight curve. This is represented as $v_{p_{i-1}, p_{i}}/\sigma$.

\begin{figure*}[!ht]
\centering
\includegraphics[width=1\linewidth]{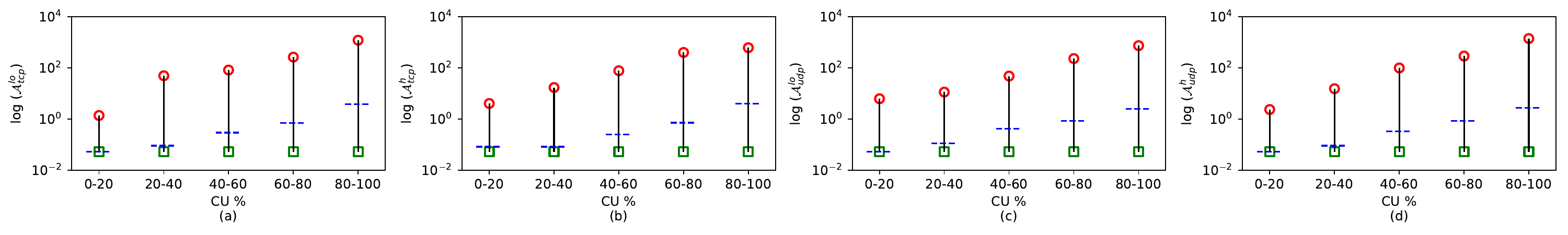}
\caption{Illustration of different accumulation scores for various \gls{cu} ranges and probe packets. 
Each sub-plot represents a specific accumulation score: (a) $ \mathcal{A}_{\text{tcp}}^{\text{lo}}$, (b) $\mathcal{A}_{\text{tcp}}^{\text{h}}$, 
(c) $ \mathcal{A}_{\text{udp}}^{\text{lo}}$, and 
(d) $ \mathcal{A}_{\text{udp}}^{\text{h}}$. 
The y-axis is log-scaled to facilitate a better understanding of the relationship between accumulation scores and the ranges of \gls{cu}.
Circles, squares, and horizontal lines represent maximum, minimum, and median values, respectively.
These results confirm the direct relationship between the accumulation score and \gls{cu}.
}
\label{as_ecdf}
\end{figure*}

\subsection{Accumulation Score for Various Probe Packet Types}

As discussed earlier, various probe packets trigger different packet processing paths of devices.
For instance, a TCP-SYN packet requires more processing compared to a UDP packet.
Also, various packet sizes induce different overheads in terms of reception processing.
For instance, the reception of a larger packet requires a longer reception time due to longer \gls{nic} to memory transfer and buffer management.
Based on these observations, we use device latency and accumulation scores for the four probe packets explained in Section \ref{mot_dl}.
Table \ref{proposed_features} summarizes the device latency and accumulation score features we use to train the \gls{ml} algorithms.

\begin{table}[!tb] 
 \centering
 \footnotesize
\def\arraystretch{1.25}
    \caption{Proposed Features for device identification}
        \begin{tabular} {|p{.9cm}|p{6.5cm}|}
        \hline
              Symbol                            &  \hfil Definition             \\ \hline  \hline
             \hfil$l_{tcp}^{lo}$                          &  Device latency for low-payload (0 B) TCP-SYN probe \\ \hline
             \hfil $\mathcal{A}_{tcp}^{lo}$                & Accumulation score for low-payload (0 B) TCP-SYN probe    \\ \hline
             \hfil $l_{tcp}^{h}$                          &   Device latency for high-payload (1400 B) TCP-SYN probe       \\ \hline
             \hfil $\mathcal{A}_{tcp}^{h}$                &   Accumulation score for high-payload (1400 B) TCP-SYN probe    \\ \hline
             \hfil $l_{udp}^{lo}$                          & Device latency for low-payload (0 B) UDP probe                       \\ \hline
             \hfil $\mathcal{A}_{udp}^{lo}$                &  Accumulation score for low-payload (0 B) UDP probe        \\ \hline
             \hfil $l_{udp}^{h}$                           & Device latency for high-payload (1400 B) UDP probe                         \\ \hline
             \hfil $\mathcal{A}_{udp}^{h}$                &   Accumulation score for high-payload (1400 B) UDP probe       \\ \hline
        
\end{tabular}
\label{proposed_features}
\end{table}

Following data collection, we determine the parameters of the bell and gamma functions for each of the aforementioned packet types to maximize the Pearson correlation between device latency and accumulation score.
For instance, we adjust the gamma function's shape and scale parameters.
We found correlation values of 92\% and 87\% for the Bell and Gamma weight function, respectively.
Consequently, we employ the Bell curve to determine the accumulation score considering its higher correlation.

\subsection{Effectiveness of Representing CU via Accumulation Scores}

To validate the effectiveness of the accumulation score as a tool for modeling \gls{cu}, we conduct an analysis estimating the accumulation score of various probe packets across different \gls{cu} ranges.
The details of data collection for this analysis can be found in Section \ref{result}.
The outcomes of this analysis are depicted in Figures~\ref{as_ecdf} (a) to (d).
A key observation from Figure~\ref{as_ecdf} is the trend of increasing accumulation score values directly related to \gls{cu} levels. 
This trend is consistent across all four types of probe packets. 
Specifically, we note that lower \gls{cu} ranges, particularly from 0 to 20\%, are characterized by lower accumulation scores, while the highest \gls{cu} range (80 to 100\%) corresponds with the maximum accumulation scores. 
This pattern indicates a positive correlation between the accumulation score and \gls{cu}, suggesting that higher channel utilization is associated with higher accumulation scores.
Further reinforcing this observation, our regression analysis demonstrates that an exponential function captures the relationship between the accumulation score and \gls{cu}. 
This is evidenced by a \gls{RSS} value of less than 0.004, indicating a strong fit.

Although a consistent increase in accumulation scores can be observed with the increase of \gls{cu} for all four probe packet types employed in this work, there are overlapping values of accumulation scores for different \gls{cu} ranges. 
This observation is due to the highly-dynamic nature of \gls{cu} and accumulation score.
Such randomness can result in overlaps of collected data instances. 
While we strive to control the \gls{cu} values by generating background traffic (exchanged between Machines 1, 2, and 3), such control is still very rough and cannot ensure that the instantaneous \gls{cu} values are always accurately or stably generated.

\begin{figure}[!t]
\centering
\includegraphics[width=.9\linewidth]{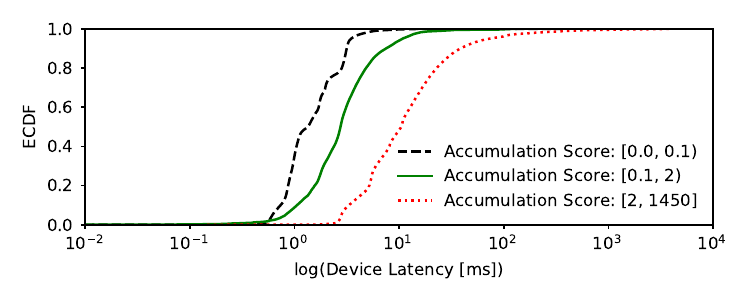}
\caption{Device latency ($l$) distribution across all devices and various accumulation score ranges.
With the increase in accumulation score, the device latency also increases.}
\label{di_rtt_as}
\end{figure}

Figure~\ref{di_rtt_as} shows the distribution of device latency for various ranges of accumulation score values and all the IoT devices in our testbed.
We adopt three bins of varying lengths to account for the fluctuations observed in accumulation scores and the presence of a long tail in the accumulation score distribution. 
It is evident from Figure~\ref{di_rtt_as} that an increase in the accumulation score corresponds to a rise in device latency. 
This further validates the effectiveness of using accumulation scores to represent \gls{cu}.

\section{Empirical Evaluations and Discussion}  \label{result}

In this section, we demonstrate the importance of using accumulation scores in data collection and model training.
Also, we study the impact of using various subsets of features on accuracy and reveal the effectiveness of the proposed features and various \gls{ml} algorithms in identifying devices.

% choice of machine learning algorithm
\subsection{\gls{ml} Algorithms}
\label{ml_alog}

\gls{ml} has emerged as a promising approach for efficiently identifying devices~\cite{9492269}.
\gls{ml} algorithms can extract data patterns effectively, facilitating device identification. 
As mentioned earlier, we choose the parameters of the weight functions to maximize the correlation between device latency and accumulation scores.
Consequently, when selecting \gls{ml} algorithms that leverage correlated features as input variables, it is essential to ensure the efficacy of these algorithms in handling such interrelated features. 
This consideration holds significance due to the possible introduction of redundancy and correlation of features, potentially introducing biases into the algorithm's predictive outcomes. 
Moreover, managing correlated features might complicate separating each feature's distinct impact on the intended target.
With this, our objective is to explore \gls{ml} algorithms that can effectively address the relationship between device latency and accumulation scores. 
Additionally, when selecting an appropriate \gls{ml} algorithm to train and deploy on residential \gls{ap}s, it is essential to consider the limited processing power and memory of \gls{ap}s~\cite{doshi2018machine}. 
Therefore, in this paper, we study tree-based algorithms such as \gls{dt} and \gls{rf}, as well as more advanced algorithms such as \gls{lgbm} and \gls{XGBoost}.

\subsection{Data Collection and Preprocessing}

In this section, we outline the design of feature rows for \gls{ml} algorithms, which are constructed to encapsulate device latency related to specific probe packets alongside their corresponding accumulation scores. The following feature vector represents this formulation:

\begin{equation}
\mathbf{x} = [ l_{\text{tcp}}^{\text{lo}}, \mathcal{A}_{\text{tcp}}^{\text{lo}}, 
                l_{\text{tcp}}^{\text{h}}, \mathcal{A}_{\text{tcp}}^{\text{h}}, 
                l_{\text{udp}}^{\text{lo}}, \mathcal{A}_{\text{udp}}^{\text{lo}}, 
                l_{\text{udp}}^{\text{h}}, \mathcal{A}_{\text{udp}}^{\text{h}}]
\label{feat_equ}
\end{equation}

Utilizing the testbed detailed in Section \ref{prpo_feature}, we have collected a comprehensive dataset that includes packet captures and device latency measurements. 
The dataset encompasses 125,336 device latency samples, with approximately 25,000 samples for each device.
To ensure a diverse representation of \gls{cu} values, background traffic has been generated to cover the possible range of \gls{cu}.
Also, as we will show in Section \ref{imp_as_dev_identf}, we map \gls{cu} intervals to accumulation score intervals and ensure that a balanced number of samples are collected for each accumulation score interval and each device.

We employ the scikit-learn Python library~\cite{scikit-learn} to implement \gls{ml} algorithms and validate the proposed features' efficacy.
% Default hyperparameters are used for the \gls{ml} algorithms.
%
Furthermore, we have addressed the issue of class imbalance in preprocessing.
Given that device identification is framed as a classification problem, a class imbalance arises when there is a disparity in the number of samples across different classes.
In particular, we need to ensure that the number of samples used for devices and \gls{cu} ranges are balanced while training or testing the models.
To mitigate this, we employ a stratification technique to maintain balanced class distributions.

\subsection{Importance of Accumulation Score and Channel Utilization for Device Identification}
\label{imp_as_dev_identf}

\begin{figure}
    \centering
    \includegraphics[width=0.7\linewidth]{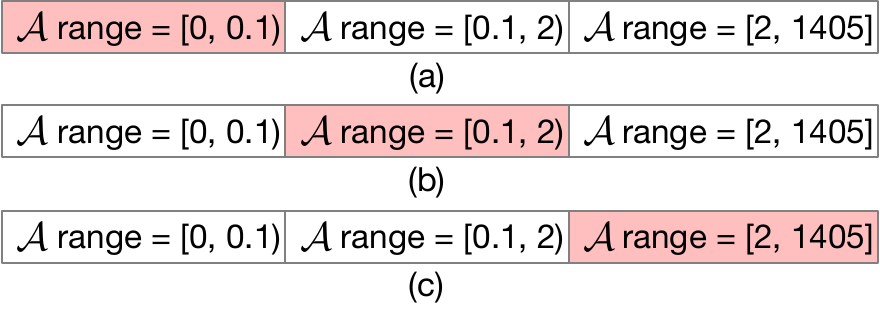}
    \caption{The training (highlighted boxes) and testing scenarios (white boxes) correspond to the results presented in Figure~\ref{acc_acc_score}.
    For instance, sub-figure (a) here corresponds to sub-figure (a) of Figure~\ref{acc_acc_score}, demonstrating the results when the LGBM algorithm is trained with samples with $\mathcal{A}$ in range [0, 0.1) and tested on samples with $\mathcal{A}$ in range [0.1, 2) and [2, 1405].
    }
    \label{as_range_testing}
\end{figure}

To demonstrate the influence of \gls{cu} on device identification accuracy, we use \gls{lgbm} algorithm and train the model within predefined accumulation score ranges and subsequently assess the performance of device identification across different accumulation score ranges. 
The details of these accumulation score ranges, designated for both training and testing, are illustrated in Figures~\ref{as_range_testing} (a) to (c).
Each defined accumulation score range bin comprises approximately 11,000 samples. 
We adopt a progressive training approach in our analysis, where we start the training process with an initial set of 1,000 samples and then gradually increase the training dataset size, adding samples incrementally until reaching 11,000.
Throughout this process, we maintain a constant number of test samples at 1,000 for each iteration.

The division of the accumulation score range is structured into three distinct intervals: $[0, 0.1)$, $[0.1, 2)$, and $[2, 1405)$. 
These intervals are approximately mapped to the corresponding \gls{cu} ranges of $[0, 33\%)$, $[33\%, 66\%)$, and $[66\%, 100\%)$. 
This mapping is established through the collection of accumulation score samples under varying \gls{cu} intensities, and then employing a regression analysis to correlate the accumulation scores with \gls{cu}. 
The \gls{cu} range is then segmented into three intervals based on this correlation, with each interval's starting and ending points derived from the corresponding segments of the regression line.

\begin{figure}[!t]
\centering
\includegraphics[width=1\linewidth]{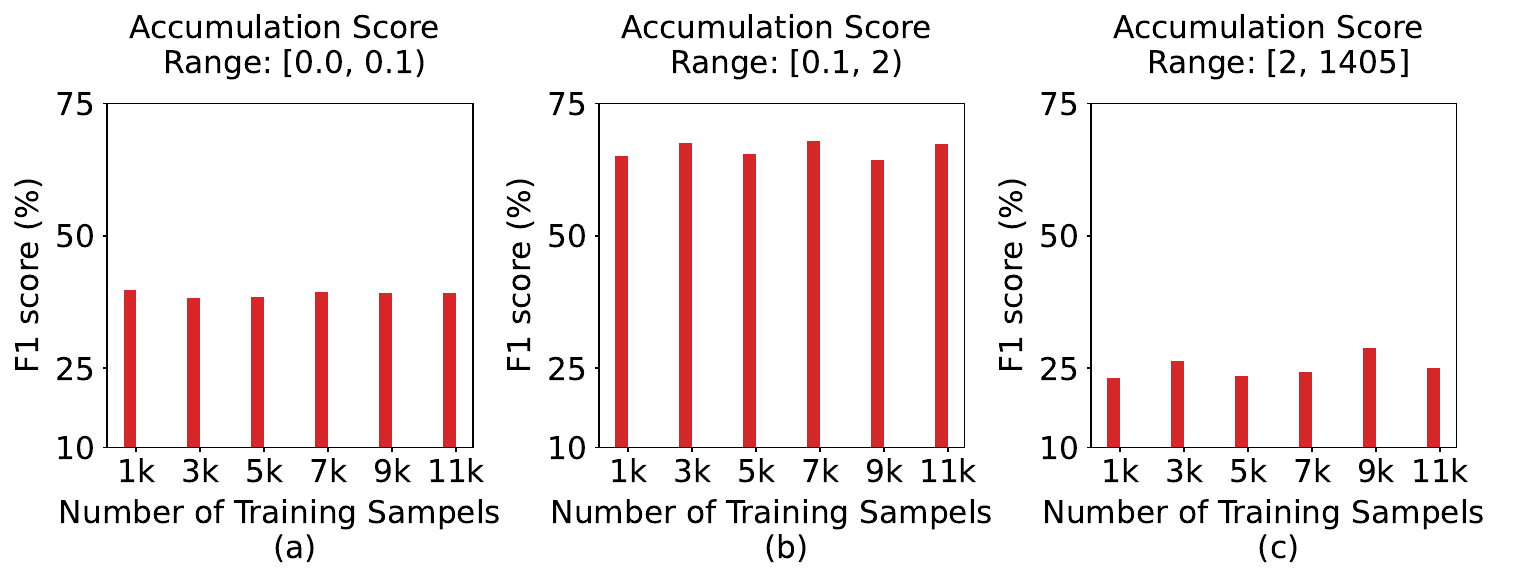}
\caption{
F1 score of device identification accuracy when the LGBM algorithm is trained using the samples belonging to one accumulation score range but tested with samples belonging to the other two accumulation score ranges.
The low F1 scores highlight the necessity of involving a diverse range of accumulation scores in the training data.}
\label{acc_acc_score}
\end{figure}

Figures~\ref{acc_acc_score} (a) to (c) present the variation of the F1 score as a function of different training sample sizes, with the x-axis depicting the number of training samples used. 
The analysis reveals a notable trend: when the algorithm is trained exclusively within a specific accumulation score range, the F1 score consistently falls below 75\%, sometimes even dropping to as low as 25\%. 
This pattern persists despite increases in the size of the training dataset, with F1 scores remaining under the 75\% threshold.
This finding underscores an important limitation: algorithms trained on data from a singular \gls{cu} range exhibit poor generalization when applied to device identification tasks across different \gls{cu} ranges. 
The model effectively learns and recognizes patterns only within the narrow confines of its training range; consequently, when confronted with \gls{cu} levels outside this familiar range, the model's ability to accurately identify devices diminishes.
Merely increasing the training dataset within the same accumulation score range does not rectify this issue, as it does not expose the model to the broader spectrum of patterns associated with different \gls{cu} levels.
We also observe a relative improvement in accuracy when the algorithm is trained using the middle accumulation score interval (i.e., $[0.1, 2)$). 
We attribute this enhanced performance to the overlap of \gls{cu} samples from lower and upper intervals within the middle range. 
As a result, the model gains partial exposure to device behaviors across the broader spectrum of \gls{cu} ranges, and this exposure enables the algorithm to partially understand device patterns outside its primary training interval.

\begin{figure}[!t]
\centering
\includegraphics[width=1\linewidth]{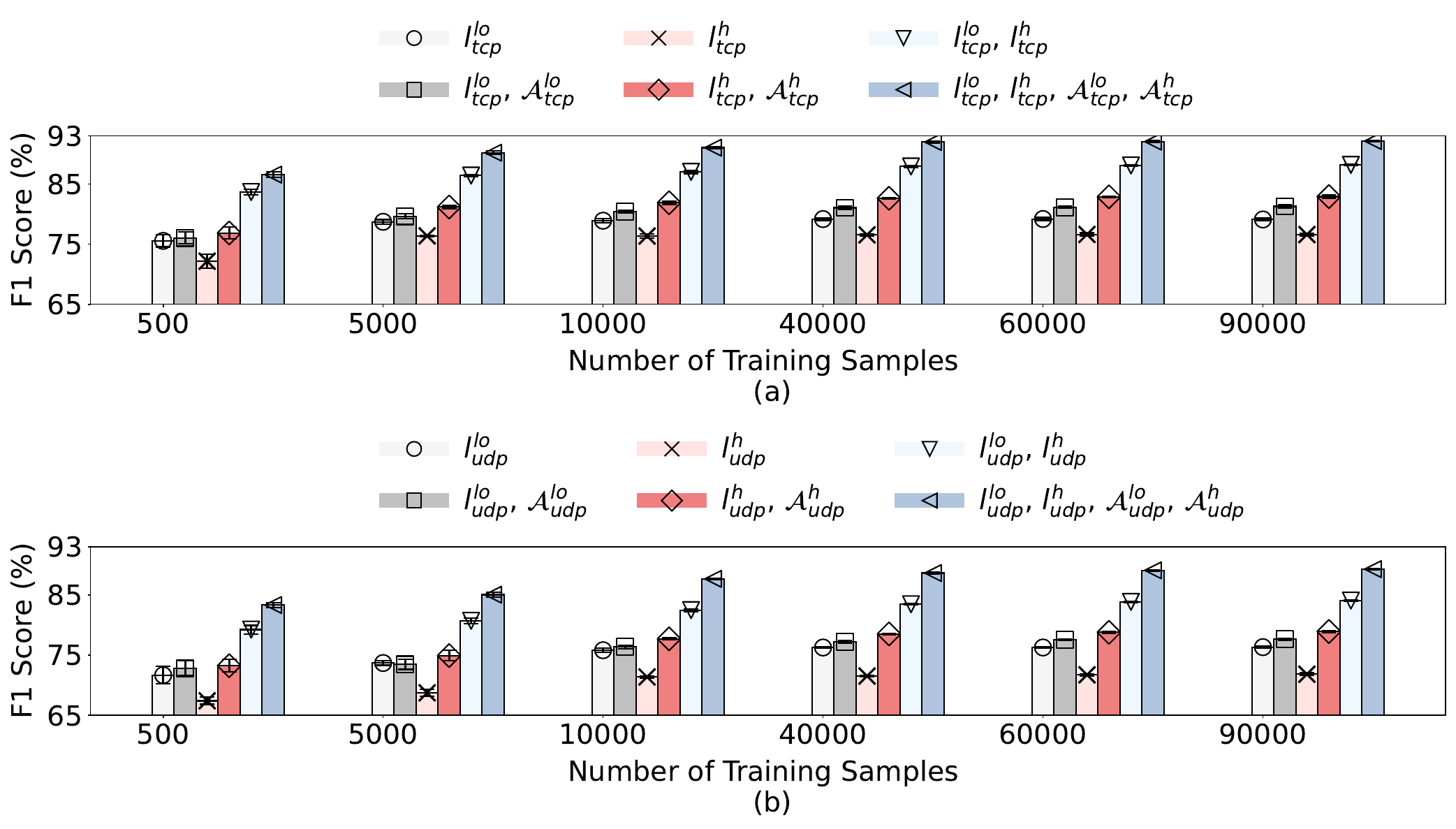}
\caption{
The F1 scores are achieved by the LGBM algorithm, with each score corresponding to the model trained on different subsets of features. 
The legend's columns represent various combinations of device latency features and their associated accumulation score features. 
%To elaborate, for a specific number of training samples, the first bar in the graph depicts the F1 score when only the $l^{lo}_{tcp}$ feature is used for training, and the second bar indicates the F1 score when the model is trained using both the $l^{lo}_{tcp}$ feature and its corresponding accumulation score feature, $\mathcal{A}^{lo}_{tcp}$. 
Error bars represent standard deviation for 30 iterations.
}
\label{fig:ch_acc_score_2}
\end{figure}

\begin{figure}[!t]
\centering
\includegraphics[width=.9\linewidth]{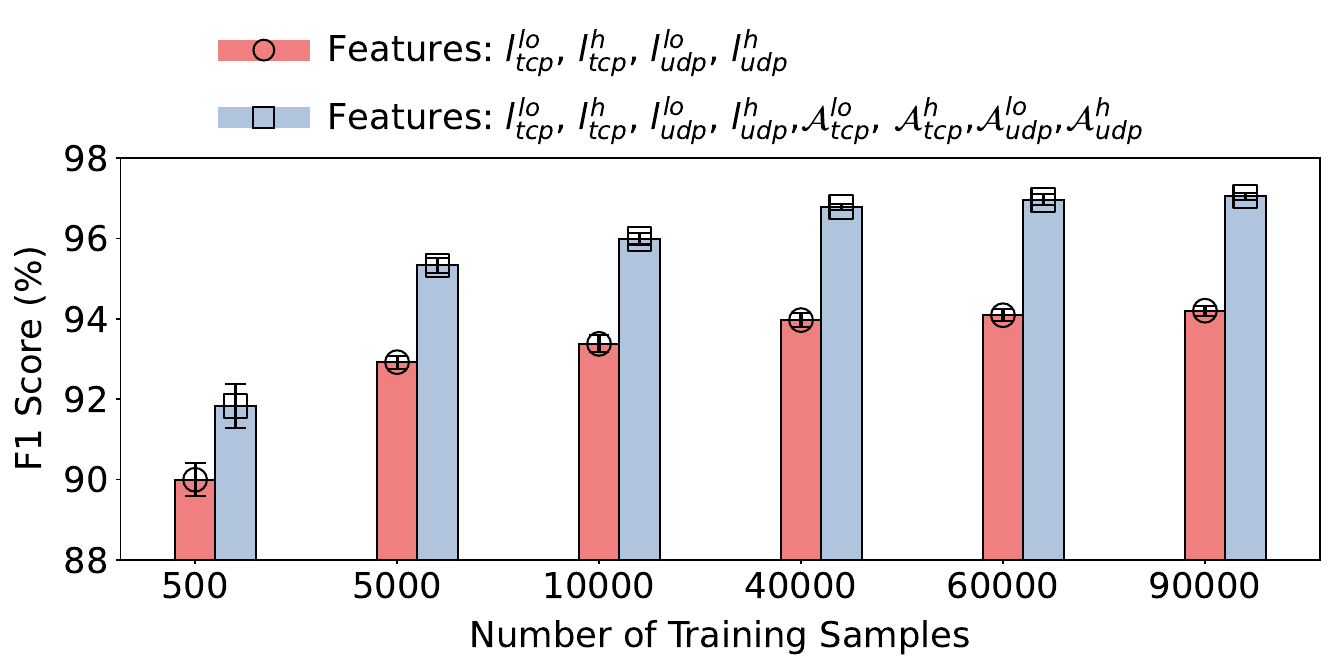}
\caption{Device identification F1 score of LGBM algorithm when using latency features (circles) and when using both latency and accumulation score features (squares).
Error bars represent standard deviation for 30 iterations.
}
\label{fig:ch_acc_score_1}
\vspace{-3mm}
\end{figure}

To further validate the significance of utilizing accumulation score for device identification, we assess the performance of \gls{lgbm} algorithm across various feature sets and present the results in
Figures~\ref{fig:ch_acc_score_2} and \ref{fig:ch_acc_score_1}.
We utilize a randomized dataset partitioning method to evaluate the algorithm's robustness and mitigate the effects of potential data variability on performance. 
This procedure is repeated by increasing the number of samples.
We iterate this process 30 times for each sample size.

Compared to Figure~\ref{acc_acc_score}, the results presented in Figures~\ref{fig:ch_acc_score_2} and \ref{fig:ch_acc_score_1} represent higher accuracy when the model has been trained with samples spanning the complete range of accumulation scores (and \glspl{cu}).
Therefore, as expected, the F1 scores are considerably higher than those presented in Figure~\ref{acc_acc_score}.
For instance, Figure~\ref{acc_acc_score} (c) shows F1 scores of around 25\% when all the latency and accumulation score features are used, whereas Figure~\ref{fig:ch_acc_score_1} shows F1 scores above 65\% even when only one feature is used.
Therefore, these results confirm the importance of training model across the range of accumulation scores.

The results in Figures~\ref{fig:ch_acc_score_2} and \ref{fig:ch_acc_score_1} confirm that even when the model has been trained across the range of accumulation scores, including the accumulation score as a feature consistently yields improved F1 scores.
Also, the accumulation score allows us to achieve higher F1 scores when the usage of some of the latency features is restricted, or the number of samples is limited.
For instance, consider a scenario where only UDP features can be used instead of TCP. 
Such restrictions may be enforced for reasons such as the lower overhead of UDP on devices' resource consumption compared to TCP, or the restrictions on sending non-legitimate, crafted TCP-SYN packets in a network due to security or firewall configurations.
In this case, the extraction and use of associated accumulation scores features with UDP latency features result in considerable performance improvement.
For instance, when we have 500 samples, Figure~\ref{fig:ch_acc_score_2}(b) shows that using feature vector $\mathbf{x} = [l^{lo}_{udp}, \mathcal{A}^{lo}_{udp} ]$ instead of feature vector $\mathbf{x} = [\mathcal{A}^{lo}_{udp} ]$ increases the F1 score from 67\% to 73\%.
It is important to note that the calculation of the accumulation score for each latency feature is implicit and does not impose any additional data exchange overhead.

\subsection{Performance Validation for Different ML Algorithms} \label{ml_algo}

\begin{figure}[!t]
\centering
\includegraphics[width= \linewidth]{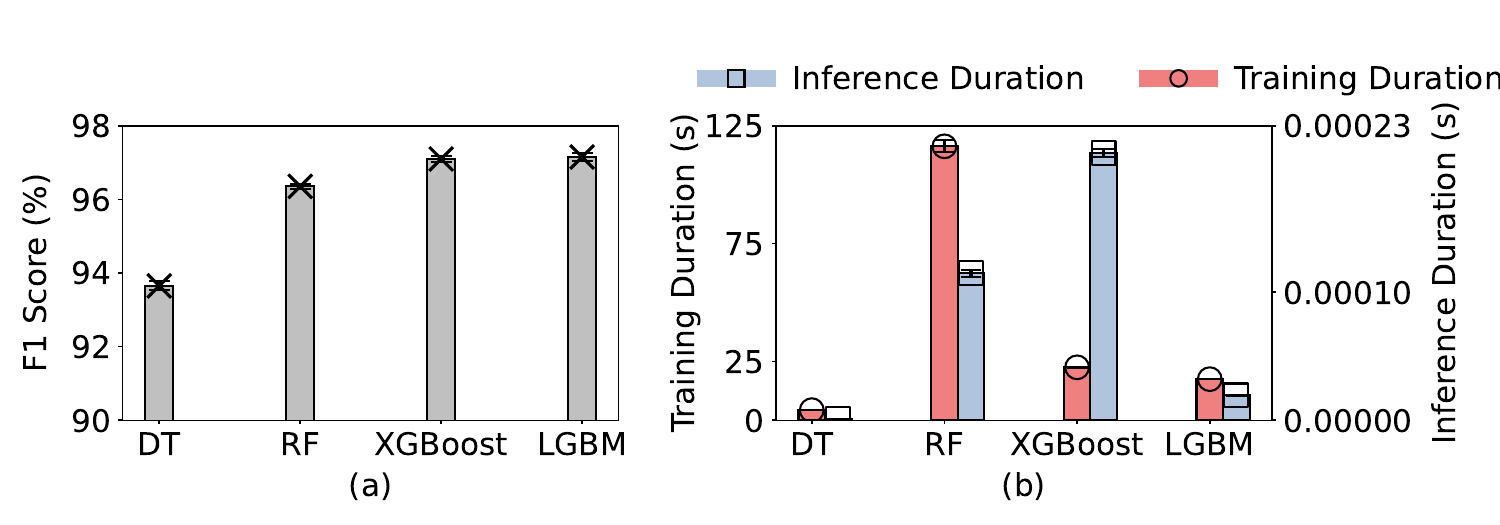}
\caption{ (a): F1 score of various \gls{ml} algorithms.
(b): Duration of training (circles) and inference (square).
Error bars represent standard deviation for 30 iterations.
}
\label{class_report}
\vspace{-3mm}
\end{figure}

In this section, we study the accuracy and overhead of various \gls{ml} algorithms.
We employ random dataset partitioning to understand the impact of training data variations on model performance and robustness.
Specifically, we utilize the whole dataset, including 125,336 samples, and partition it into distinct training and testing sets with an 80:20 split ratio. 
This process is iterated 30 times.

Figure~\ref{class_report}(a) shows the accuracy of various \gls{ml} algorithms.
\gls{dt} demonstrates the lowest F1 score of 93.6\%, and \gls{lgbm} (and \gls{XGBoost}) achieves the highest score of around 97\%. 
\gls{dt}s are characterized by their simplicity and ease of interpretation; however, they have certain limitations. 
The implementation of \gls{dt}s in scikit-learn library adopts binary splits in features, forming a hierarchical structure. 
This procedure generates a decision sequence that separates data subsets resembling the target class. Since these decisions depend on a succession of binary divisions, they might not be able to capture complex relationships among multiple features within datasets. 
\gls{rf} employs an ensemble approach involving multiple \glspl{dt}.
In this approach, each \gls{dt} is trained on a separate random subset of training data and features, introducing diversity among the individual trees. 
\gls{rf} determines the prediction outcome by aggregating the predictions from \glspl{dt}. 
This ensemble methodology enhances the F1 score compared to a \gls{dt}. 
Finally, \gls{lgbm} and \gls{XGBoost} are ensemble algorithms that utilize gradient-boosting techniques.
They combine the predictions of weak learners (such as \gls{dt}) such that they focus on misclassified samples.
Leveraging gradient descent techniques, \gls{lgbm} and \gls{XGBoost} iteratively optimize the algorithm performance over multiple boosting rounds.
This approach enables these algorithms to handle complex and nonlinear relationships effectively and achieve improved F1 scores, outperforming the standalone \gls{dt} and \gls{rf}.
Figure~\ref{class_report}(a) also shows the low standard deviation of F1 scores, indicating low bias and variances in these \gls{ml} algorithms. 
It further suggests that the proposed features are robust and stable across different ML algorithms and not influenced by the dataset's random partitioning.
These observations indicate that utilizing the device latency and accumulation score as features has significant promise in providing reliable and stable results in practical applications.

The Netgear WAX 218 \gls{ap} includes an IPQ8074 SoC~\cite{IPQ8074}, which integrates a quad-core Cortex-A53 processor.
Using this processor, we measure the duration of training and inference and present the results in Figure~\ref{class_report}(b).
Note that inference duration refers to the duration of making an inference using the feature vector presented in Equation \ref{feat_equ}.
The results show that \gls{dt} is the fastest model to train and has the shortest inference duration, suggesting a suitable solution for resource-constrained scenarios. 
Although the computational overhead of \gls{dt} is the lowest, its accuracy is also the lowest, providing a possible trade-off between computational efficiency and classification performance.
\gls{rf} inherently requires more time for training (116.27 seconds) due to the complexity of constructing multiple trees. 
Nevertheless, its inference time per sample remains relatively small (0.000114 seconds), benefiting from the ensemble's majority voting mechanism during prediction.
Given the similar F1 scores achieved by \gls{lgbm} and \gls{XGBoost}, the algorithm with lower overhead better fits the resource-constrained nature of \glspl{ap}.
We observe that the training and inference duration of \gls{lgbm} are 5 seconds and 0.0001893 seconds, respectively, shorter than those of \gls{XGBoost}. 
This is because \gls{lgbm} significantly reduces memory overhead by representing data distributions using histograms.
This reduction in memory overhead results in minimized memory consumption for processing and storing intermediate results, potentially reducing training and inference duration~\cite{ke2017lightgbm}. 
The use of \gls{lgbm} is specifically beneficial for resource-constrained devices, such as home \glspl{ap}.

\section{Related Work} 
\label{realted_work}

Recent works on IoT device identification center around leveraging traffic-based features to achieve high accuracy. 
We have classified these studies into a few distinct categories, determined by the types of features they utilize.

The first category of features is extracted by analyzing individual network packets. 
For example,~\cite{valdez2019discover} utilized TLS protocol features, such as TLS version, server name, IP address, and negotiated cipher suites, for device identification. 
Similarly, the study in~\cite{8761559} adopted features like UDP checksum, TCP window size, DNS query class, IP length, TCP port, and TCP flags, focusing on packet headers and network traffic.
Expanding the scope,~\cite{miettinen2017iot} extracted features across 16 protocols, including HTTP, DNS, and NTP, analyzing packet size, port numbers, and IP address counts for each protocol type. 
Specifically, instead of aggregating traffic statistics over time, they built feature vectors representing the features of a specific number of packets.
Moreover,~\cite{thangavelu2018deft} focused on features such as packet length and the count of specific packet types, including STUN, NTP, and MQTT.
While these studies achieve high accuracy (above 90\%) in device identification, their reliance on analyzing packet contents raises privacy and computational resource concerns. 
Sensitive packet contents, like DNS requests, device IP addresses, and port numbers, may pose privacy issues for users. 
Also, stringent data protection regulations could further restrict access to such packet contents, complicating the analysis. 
Additionally, parsing packet contents, especially payloads, often requires significant computational and memory resources. 
% challenging the use of 

The second category of features emphasizes flow-based statistics aggregated across packets, as opposed to focusing on individual packet details. For example,~\cite{gordon2021securing} proposed flow-based features like the percentages of ICMP, TCP, and UDP protocols, packet count and size, and the diversity ratio of IP addresses.
In~\cite{shaikh2018machine}, the authors utilized metrics such as average time-to-live per IP address, total packet count, unique IP addresses and port counts, and TCP/UDP packet numbers. 
The authors in~\cite{cvitic2021ensemble} focused on packet lengths, both sent and received, and the maximum length of sent packets. 
Features like packet and byte counts, mean packet size, inter-arrival time, and the count of TCP packets with specific flags were used in~\cite{salman2022machine}.
This approach was refined in~\cite{santos2018efficient} and~\cite{pinheiro2019identifying}, where the former used packet size, inter-arrival time, and specific TCP flag counts, while the latter tailored their analysis to encrypted traffic, examining packet length statistics over a one-second window.
In~\cite{fakhruldeen2024enhancing}, the study identified IoT devices in Wi-Fi environments, emphasizing data frame feature selection from the 802.11 link layer. 
They extracted eleven key features, including frame length, type, arrival time, and retransmission flag. 
Despite the strengths of flow-based features, gathering comprehensive training data to represent flow statistics accurately can be challenging and time-consuming, especially for IoT devices that operate intermittently and under resource constraints. Furthermore, these flow-based features can be sensitive to network conditions, significantly limiting their generality across different network environments.  
For instance, as demonstrated in this paper, high \gls{cu} during peak times can alter flow patterns, impacting identification accuracy. 

Several studies have augmented flow-based features with packet-specific features to leverage the benefits from both categories to improve identification accuracy further. 
For example,~\cite{ammar2020autonomous} combined flow-based features like inter-arrival times and packet counts with packet-specific ones across various protocols. 
Similarly,~\cite{ullah2021network} and~\cite{hamad2019iot} demonstrated the effectiveness of merging packet features with flow-based statistical properties, including source ports, protocols, and packet rates. In~\cite{sivanathan2018classifying}, the authors examined both individual packet features, such as domain names, cipher suites, DNS intervals, and port numbers, as well as flow-level features, like the total number of bytes downloaded/uploaded, the duration of flows, and the inactive intervals of IoT devices. 
%both individual packet and flow-based features. 
%
Although these studies lead to continuous improvement in identification accuracy, there is a lack of solutions that can address privacy concerns and features' reliance on network conditions at the same time.

To overcome these challenges, this work introduces a new feature type -- accumulation score -- which captures instantaneous \gls{cu} and its impact on device latency features. Experiment results validate that the device identification scheme integrating accumulation score with the proposed latency features can protect user privacy while achieving high and robust accuracy across different wireless channel dynamics. In addition, this work also proposes a proactive and expedited data collection method that captures device latency in response to different packet types. 

\section{Conclusion} 
\label{conc}
In this paper, we focused on the accuracy of IoT device identification, taking into account two critical design factors. 
Firstly, we utilized device latency instead of deep packet inspection methods due to the latter's high computational demands and privacy concerns. 
Secondly, we highlighted the significant influence of wireless channel dynamics on device latency features, underscoring the necessity of incorporating these dynamics into the device identification process. 
To effectively model channel dynamics, we introduced the accumulation score as a metric that is derived from packet capture data and accounts for the impact of wireless channel utilization on the variability of features used in device identification. 
Our findings reveal that the accumulation score facilitates enhanced data collection for training and also significantly improves the accuracy of device identification when incorporated as an additional feature in training and testing \gls{ml} algorithms.

\section*{Acknowledgement}
We extend our gratitude to Netgear Inc. (San Jose, USA) for their generous donation of devices crucial for the execution of this research.

\bibliographystyle{IEEEtran}
\bibliography{IEEEfull.bib}

\balance

\end{document}